\documentclass[final,5p,times,twocolumn,numbers]{elsarticle}

\usepackage{hyperref}
\usepackage{lipsum}
\usepackage[utf8]{inputenc}
\usepackage[version=4]{mhchem}
\usepackage{soul}
\usepackage{siunitx}
\usepackage{makecell}
\usepackage{multirow}
\usepackage{amsmath,amssymb}
\usepackage{braket}
\usepackage{float}
\usepackage{subcaption}
\usepackage{url}
\usepackage{tabularx}

\usepackage{cleveref}
\crefname{figure}{Fig.}{Figs.}
\Crefname{figure}{Fig.}{Figs.}
\crefname{section}{Section}{Sections}
\Crefname{section}{Section}{Sections}
\crefname{table}{Table}{Tables}
\Crefname{table}{Table}{Tables}
\crefname{appendix}{Appendix}{Appendices}
\Crefname{appendix}{Appendix}{Appendices}
\crefname{subfigure}{Figure}{Figures}
\Crefname{subfigure}{Figure}{Figures}

\newcommand{\overbar}[1]{\mkern 1.5mu\overline{\mkern-1.5mu#1\mkern-1.5mu}\mkern 1.5mu}

\biboptions{sort&compress}

\journal{Physics Letters B}

\begin{document}

\begin{frontmatter}



\title{Chemically-polarized material for nuclear and particle physics }


\author[inst1,inst2]{Benjamin G. Collins}
\author[inst1]{Daniel P. Watts\corref{cor1}}
\author[inst1]{Mikhail Bashkanov}
\author[inst1]{Stephen Kay}
\author[inst2]{Simon B. Duckett}
\author[inst3]{Andreas Thomas}
\author[inst4,inst5,inst6,inst7]{Dmitry Budker}
\author[inst8]{Danila Barskiy}
\author[inst4,inst5,inst6]{Raphael Kircher}

\cortext[cor1]{Corresponding author. Email: daniel.watts@york.ac.uk.}

\affiliation[inst1]{
organization={Department of Physics, Engineering and Technology, University of York},
city={Heslington},
postcode={YO10 5DD}, 
country={UK}}

\affiliation[inst2]{organization={Centre for Hyperpolarisation in Magnetic Resonance, University of York},
city={Heslington},
postcode={YO10 5NY}, 
country={UK}}

\affiliation[inst3]{
organization={Institut für Kernphysik, Universität Mainz},
city={Mainz},
postcode={55128}, 
country={Germany}}

\affiliation[inst4]{
organization={Johannes Gutenberg Universität},
city={Mainz},
postcode={55128}, 
country={Germany}}

\affiliation[inst5]{
organization={Helmholtz-Institut Mainz},
city={Mainz},
postcode={55128}, 
country={Germany}}

\affiliation[inst6]{
organization={GSI Helmholtzzentrum für Schwerionenforschung GmbH},
city={Darmstadt},
postcode={64291}, 
country={Germany}}

\affiliation[inst7]{
organization={Department of Physics, University of California},
city={Berkeley},
postcode={CA 94720}, 
country={USA}}

\affiliation[inst8]{
organization={Frost Institute for Chemistry and Molecular Sciences, Department of Chemistry, University of Miami},
city={Coral Gables},
postcode={FL 33146}, 
country={USA}}

\begin{abstract}
Spin-polarized solid targets have underpinned many recent key advances in nuclear and particle physics, yet traditional methods to produce them face significant limitations due to the high cost and demanding cryogenic and magnetic field requirements. These factors constrain experimental geometries and present challenges in intense radiation environments where depolarization and materials damage can occur. We present the first results assessing the capabilities of the chemical hyperpolarization (ChHP) method Signal Amplification By Reversible Exchange (SABRE) to act as the polarization method to produce targets or active detector media. We show by using in-beam measurements that there is no depolarizing effect observed with the SABRE-polarized material in the A2 photon beam at the Mainzer Mikrotron (MAMI), as well as showing the resilience of such media to radioactive doses of up to \SI{3}{\kilo\gray}. We also illustrate the capabilities for using SABRE-polarized material as a scintillation or Cherenkov detector.    
\end{abstract}




\end{frontmatter}

%\tableofcontents



\section{Introduction}
\label{sec:introduction}

Current solid polarized targets for nuclear and particle physics suffer from beam-induced depolarization effects when used with intense charged-particle beams, limiting the maximum target polarization and usable beam currents. These effects include beam heating \cite{liu1998depolarization,lowry2013electrons}, where the energy deposited by the beam warms the sample and increases the rate of longitudinal nuclear relaxation (characterized by the time constant $T_1$), as well as the production of radicals in the sample which act as a source of paramagnetic relaxation \cite{meyer2004ammonia,pandey2023operation}.

Ammonia (\ce{NH3}), polarized via dynamic nuclear polarization (DNP), is currently the preferred material for polarized \ce{^{1}H} targets needed to withstand high beam intensities. Its strong resistance to radical formation allows for the total accumulated dose before target replacement to be over an order of magnitude higher than for butanol or HD targets \cite{meyer2004ammonia,meyer1984dynamic}. Despite this, limitations persist; for instance, the frozen-spin \ce{NH3} target selected for Drell-Yan measurements by the COMPASS collaboration was restricted to beam currents of just \SI{16}{\pico\ampere} when used with a $\pi^-$ beam~\cite{andrieux2022large}, due to the high sensitivity of $T_1$ to heat deposition at the 60~mK frozen-spin operating temperature. Beam-induced depolarization effects are particularly restrictive for experiments utilizing frozen-spin targets, currently the only option for achieving near 4$\pi$ angular acceptance with fixed polarized targets. 

Continuously polarized targets are able to operate within more intense beams, with the typical electron beam currents used ranging from 1 to 8 nA \cite{keith2003polarized,pandey2023operation}. At this intensity, however, the targets have to be replaced every few days due to radiation damage, primarily consisting of radical build-up, and a corresponding reduction in polarization levels \cite{Keith:2023XR}. For \ce{NH_3} targets, the amine radical \ce{^{14}NH2^.} is stable for weeks at temperatures of \SI{77}{\kelvin} \cite{meyer1983irradiated}, far in excess of DNP operating temperatures, and thus this radical and others accumulate within the sample.

At storage beam facilities, polarized internal gas targets are instead frequently used as their lower areal density (${\sim} 10^{13}$ atoms \unit{\per\centi\metre\squared} \cite{Mikirtychyants:2011hov}) reduces beam degradation effects. Alternatively, for unpolarized material, a pellet target system introducing a continuous stream of small (${\sim}$\SI{10}{\micro\metre}) frozen droplets of hydrogen into the beam has been used for the WASA detector at CELSIUS/COSY \cite{ekstrom2002celsius,hoistad2004proposal}, allowing for densities of ${\sim} 10^{15}$ atoms \unit{\per\centi\metre\squared}. 

Signal Amplification By Reversible Exchange (SABRE) is a nuclear polarization technique that utilizes the high spin order of \textit{para}hydrogen (\textit{p}-\ce{H_2}), the singlet nuclear spin isomer of molecular hydrogen. SABRE is used to catalytically transfer spin order from \textit{p}-\ce{H_2} to a range of typically nitrogen-containing organic molecules including pyridines, amines, nitriles, and diazoles \cite{adams2009reversible,iali2018using,fekete2013iridium,shchepin201615n}. This spin-order transfer occurs in a transient organometallic complex, where the (typically iridium) SABRE catalyst reversibly binds the \textit{p}-\ce{H_2} and substrate molecules, facilitating spin transfer through a scalar coupling network \cite{adams2009theoretical}, spin-spin interactions transmitted through the bonding electrons. This process takes place in the solution state and, notably, can operate efficiently at room temperature. A related technique, hydrogenative PHIP, which generates polarization through the catalytic, pairwise addition of \textit{p}-\ce{H_2} across an unsaturated bond, has been previously proposed by Budker et al. as a method for producing polarized material for use in fixed-target scattering experiments \cite{BUDKER2012246}.

\begin{figure}[htbp]
	\centering 
	\includegraphics[width=0.88\columnwidth]{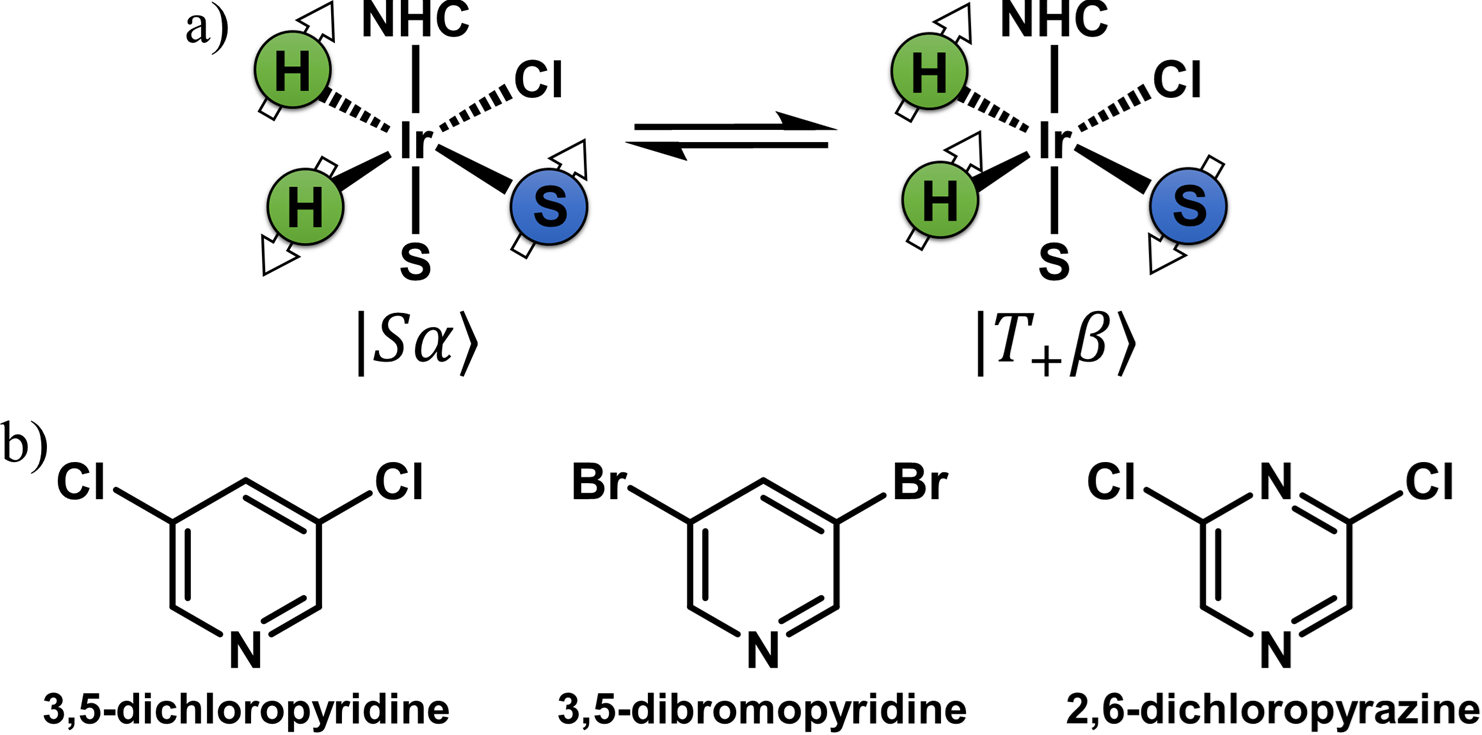}	
	\caption{a) Spin-order transfer during SABRE, converting \textit{p}-\ce{H_2} to \textit{o}-\ce{H_2} and an initially unpolarized spin to a polarized spin. The active SABRE catalyst shown here is of the form \ce{[IrCl(H)_2(NHC)(S)_2]}, where S is a bound substrate molecule and NHC = \textit{N}-heterocyclic carbene. b) Substrates investigated in this study. 
    }
	\label{fig:sabre_spin_transfer_substrates}%
\end{figure}

Coherent evolution of the spin states in SABRE takes place at a level anti-crossing (LAC), where two quantum energy levels approach each other as a function of  the magnetic field strength. The initially overpopulated state $\ket{S\alpha}$ evolves to the state $\ket{T_+\beta}$, transferring spin order from the \textit{p}-\ce{H_2}-derived iridium hydride protons to the target nucleus (\cref{fig:sabre_spin_transfer_substrates}a). Here, $\ket{S}\Rightarrow\ket{T_+}$ represents the transition from the singlet \textit{para}hydrogen state to the triplet \textit{ortho}\-hydrogen (\textit{o}-\ce{H_2}) state, and $\ket{\alpha}\Rightarrow\ket{\beta}$ represents the polarization of an initially `unpolarized' spin. This LAC condition is met at \unit{\milli\tesla} fields for \ce{^{1}H} polarization, although SABRE polarization has also been demonstrated for other nuclei including \ce{^13C}, \ce{^15N}, \ce{^19F} and more \cite{barskiy2017absence,theis2015microtesla,shchepin2017toward,burns2015improving,olaru2016using}, for which the LAC conditions are typically met at \unit{\micro\tesla}--\unit{\milli\tesla} fields.

In this work we present the first measurements of the resilience of SABRE-polarized material under particle beam exposure, utilizing the MAMI accelerator facility. This work addresses two key objectives: firstly, to measure the beam-induced depolarization by the A2 photon beam in order to assess for radical-induced $T_1$ reduction, and secondly, to determine the extent of radiation damage within the sample caused by a high irradiation dose over time, achieved in proximity to the electron beam dump in the A2 hall. Radiation damage of the samples may be expected to occur through the destruction of the SABRE catalyst and substrates through radiolysis. Radical accumulation within the sample was not expected to occur, as radical termination was expected to occur on a rapid timescale at the room temperature operating conditions of SABRE. 

This work is in reference to the clear need in the international community of nuclear and particle physics for polarized target methodologies that can operate at the frontiers of intensity. It is hoped that SABRE can address the issues facing the use of DNP targets with intense charged particle beams, benefiting from its capabilities for room-temperature operation and fast polarization build-up, achieved at low magnetic fields.

\section{Methodology}
\label{sec:methodology}

These experiments were conducted in the A2 collaboration hall, where a beam of bremsstrahlung photons was generated by the passage of the MAMI electron beam \cite{herminghaus1976design} through a thin Møller radiator (VACOFLUX 50 alloy: 49\% Co, 49\% Fe, 2\%~V). The energy of these photons was provided from the upgraded Glasgow-Mainz tagging spectrometer \cite{mcgeorge2008upgrade} which measures the energy of the scattered electrons. For these experiments, an electron beam energy of 855 MeV was used at a beam current of \SI{10}{\nano\ampere}.

The in-beam polarization decay measurements were performed for three SABRE-compatible substrates: 3,5-dichloropyridine (\textbf{3,5-dcpy}), 3,5-dibromopyridine (\textbf{3,5-dbpy}), and 2,6-dichloropyrazine (\textbf{2,6-dcpz}) (\cref{fig:sabre_spin_transfer_substrates}b), chosen for the long lifetimes of their hyperpolarized states ($T_1\approx10^2$~\unit{\second}), a necessity for this experiment. Three substrates were investigated to assess whether any potential beam-induced depolarization effects were substrate-dependent.

\begin{figure}[htbp]
	\centering 
	\includegraphics[width=0.80\columnwidth]{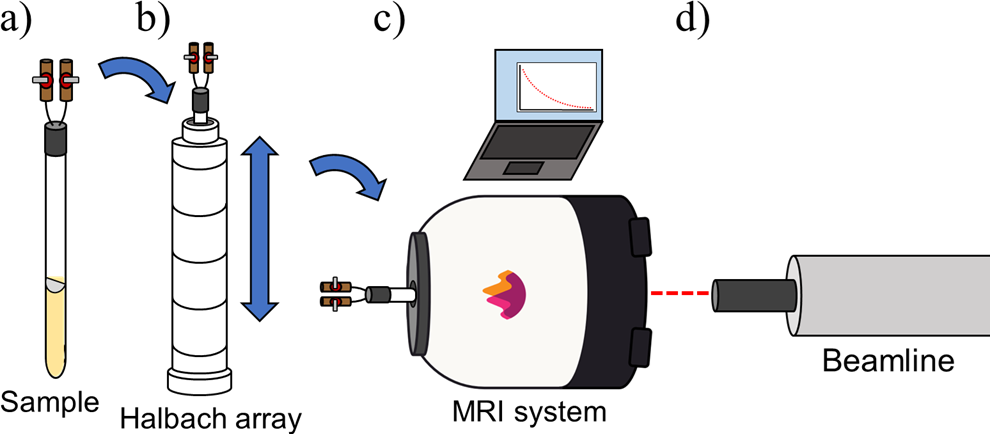}	
	\caption{Diagram of the experimental procedure. a) Prepare sample and fill with \textit{p}-\ce{H_2}. b) Transfer to Halbach array and shake for \SI{45}{\second}. c) Transfer to MRI system and start acquisition. d) Vacate hall and turn on photon beam.} 
	\label{fig:experimental_procedure}%
\end{figure}

Before each run, spent gas in the polarization cell from the previous run was removed and replenished with 5 bar of \textit{p}-\ce{H_2}. The sample was polarized by placing it within a 6 mT handheld Halbach array and shaking for 45 s to disperse the \textit{p}-\ce{H_2} throughout the solution. After polarizing, the cell was inserted horizontally into a benchtop MRI system (\SI{0.33}{\tesla}, Resonint ilumr \cite{Resonint}) that was oriented with the bore along the axis of the beamline. This MRI system was configured for NMR spectroscopy, and the decay of polarization within the sample was measured using a series of low tip-angle free induction decay (FID) scans (see supplementary material for details). Runs with the beam on were compared to control runs without the beam in order to assess for any beam-induced depolarization effects. Typical wait times between the end of the polarization process and the beam being operational were a few minutes, corresponding to $1$--$3\,T_1$ for the investigated samples. A diagram of the experimental procedure can be seen in \cref{fig:experimental_procedure}.

Throughout the four-day beamtime, a cell containing a SABRE sample was placed near the electron beam dump in the A2 hall, chosen for being the area where the highest radiation dose would be received. The relative polarization value and polarization decay rate were compared before and after irradiation to test for the effects of radiation damage on the sample. 

The investigated samples contained \SI{0.63}{\milli\mol\per\litre} of [Ir(IMes-\textit{\ce{d_22}})(COD)Cl] precatalyst, \SI{75}{\milli\mol\per\litre} of substrate, and \SI{3.1}{\milli\mol\per\litre} of dimethyl sulfoxide-\textit{\ce{d_6}}, acting as a co-ligand, dissolved in 5 mL of dichloromethane-\textit{\ce{d_2}}. This represented a 120-fold molar excess of substrate and a 5-fold molar excess of co-ligand relative to the catalyst. The samples were degassed three times using a freeze-pump-thaw technique, whereby the sample was exposed to a strong vacuum (${\approx}10^{-3}$~\unit{\milli\bar}) whilst frozen, before being thawed and purged with nitrogen. The samples were then pressurized with \textit{p}-\ce{H_2} (5 bar) and left to activate fully over \SI{3}{\hour} to form their respective active SABRE catalysts. Under similar conditions, \textbf{3,5-dcpy} and \textbf{3,5-dbpy} have previously been shown by Tickner et al.~to form the neutral active SABRE catalyst [IrCl\ce{(H)_2}(NHC)(substrate)(sulfoxide)], in contrast to the charged catalyst [Ir\ce{(H)_2}(NHC)\ce{(substrate)_2}(sulfoxide)]Cl that is seen when using more polar solvents \cite{tickner2024metal}.

\begin{figure}[htbp]
    \centering
    \includegraphics[width=1\columnwidth]{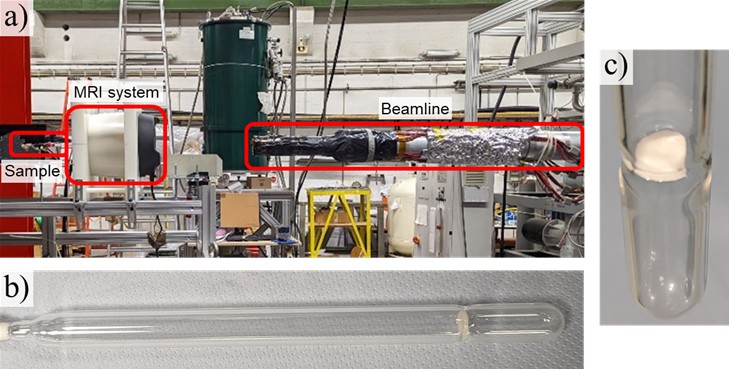}
    \caption{a) MRI system positioning next to the beamline. Here the end of the beamline and the MRI system are separated by approximately 50 cm. b) The polarization cell. c) Close-up of the glass insert in the polarization cell.}
    \label{fig:pol_cel_mri_pos_cmbined}
\end{figure}

\begin{figure}[h]
	\centering 
	\includegraphics[width=0.71\columnwidth]{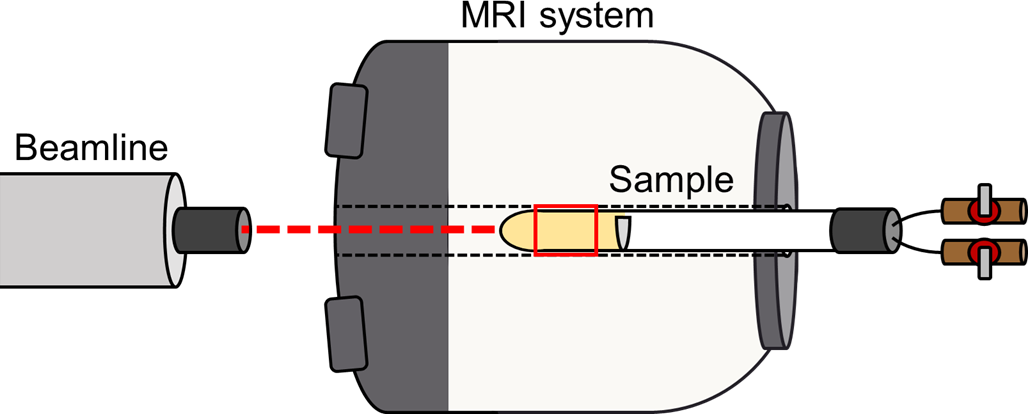}	
	\caption{Diagram showing the placement of the cell within the MRI system and in relation to the beam. The black dotted lines show the internal bore of the MRI system, the red dotted line shows the path of the beam, and the solid red lines show the sensitive region of the MRI system.} 
	\label{fig:mri_int_schem}%
\end{figure}

The experimental setup used for the in-beam measurements of the polarization decay can be seen in \cref{fig:pol_cel_mri_pos_cmbined}a. As is shown, the MRI system was positioned on its side with the axis of the bore parallel to the beamline, such that the internal bore was precisely aligned with the centre of the photon beam. A lead collimator (not shown) with an internal diameter of 3 mm and a length of 200 mm, located upstream of the MRI system, was used in order to control the cross-sectional size of the beam. The hollow bore of the MRI system allowed the photon beam to impact directly onto the bottom face of the polarization cell that is shown in \cref{fig:pol_cel_mri_pos_cmbined}b--c. The cell consisted of a 20 cm length of 15 mm OD heavy-walled borosilicate glass tubing, divided into two regions by a semicircular glass insert that obstructed the majority of the path between the two regions, leaving a narrow gap. This gap allowed for the mixing of the sample volume and \textit{p}-\ce{H_2} headspace under vigorous shaking; however, surface tension prevented liquid transfer under gravity when the cell was static. This design was necessary to ensure that the end region of the cell remained filled with the liquid sample during measurements within the MRI system (\cref{fig:mri_int_schem}). At the end of the polarization step, the sample volume was collected in the end region of the cell, before the cell was tilted horizontally with the sample remaining in the end region, held by surface tension.

\section{Results}
\label{sec:results}

\subsection{In-beam polarization decay}
\label{subsec:in-beam_pol_dec}
For the in-beam polarization decay monitoring of SABRE-polarized material in the the A2 photon beam, the total photon flux for ${\geq}$\SI{10}{\mega\electronvolt} photons at a beam current of \SI{10}{\nano\ampere} was \SI{4.2e8}{\hertz}, found by extrapolating data from the Glasgow-Mainz tagger over the range 40-855 \unit{\mega\electronvolt} using the $\frac{1}{E}$ dependence of Bremsstrahlung photons. Photons of energy less than \SI{10}{\mega\electronvolt} were effectively filtered out in this experiment by the collimator due to the larger scattering angle they exhibited. Photon flux measurements were then combined with the average energy deposition per photon, found by Geant4 simulation, giving a total rate of energy deposition of \SI{6.4e-4}{\joule\per\second}. Due to the low rate of energy deposition, we calculate that the sample experienced an average rate of beam heating of only \SI{5}{\milli\kelvin\per\minute}, which in isolation would produce no measurable effect on the $T_1$.

\begin{figure}[htbp]
    \centering
    \includegraphics[width=1\columnwidth]{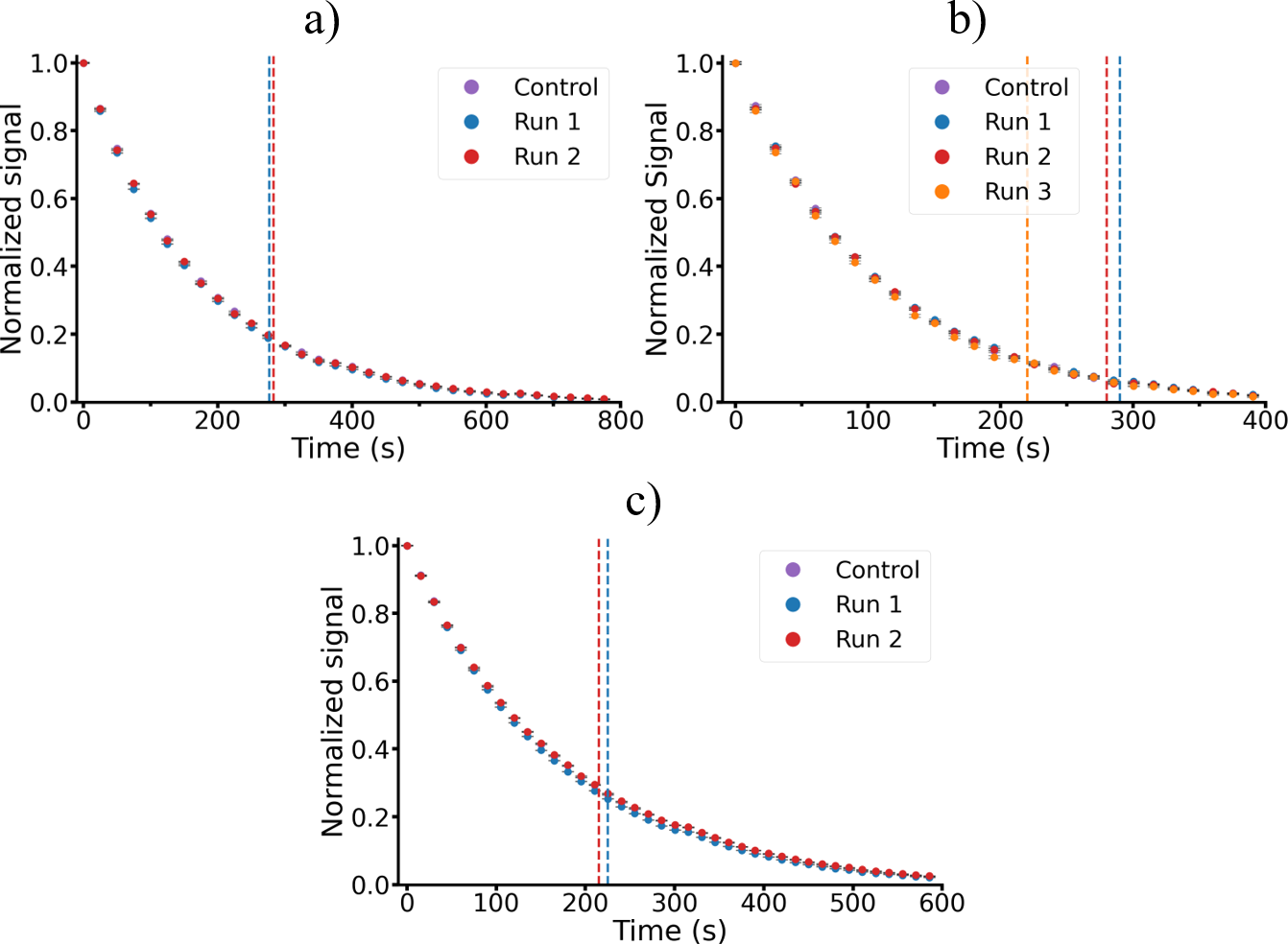}
    \caption{Normalized polarization decay profiles for a) \textbf{3,5-dcpy}, b) \textbf{3,5-dbpy}, c) \textbf{2,6-dcpz}. The dotted lines indicate the start time of the photon beam.}
    \label{fig:decay_profiles}
\end{figure}

\cref{fig:decay_profiles} shows the polarization decay profiles for all runs in the A2 photon beam, normalized to unity at $t=0$ to illustrate the relative polarization levels throughout the run. The average fitted $T_1$ values are shown in \cref{tab:decay_t1s}. The samples were measured over a period commensurate with an appreciable signal to noise ratio (SNR). The sequence of tip angles and the measurement times used for the decay profiles of each substrate can be found in the supplementary material.

\begin{table}[htbp]
    \centering
    \caption{Average $T_1$ of \ce{^{1}H} nuclear spins for all investigated substrates, for before and after the beam was turned on. For the control runs, the average start time of the beam for the beam-on runs was used.}
    %\resizebox{\columnwidth}{!}{%
    \begin{tabular}{|c|c|c|c|}
    \hline
    \multirow{2}{*}{\textbf{Substrate}} & \multirow{2}{*}{\textbf{Run}} & \multicolumn{2}{c|}{\textbf{\ce{^{1}H} \boldmath$T_1$ (s)}} \\ 
    \cline{3-4}
                                        &                               & Before & After \\
    \hline
    \hline
    \multirow{2}{*}{\textbf{3,5-dcpy}} & Control       & $170 \pm 1$ & $160 \pm 3$  \\
    \cline{2-4}
                                       & Beam-on   & $167 \pm 3$ & $160 \pm 6$  \\
    \hline
    \hline
    \multirow{2}{*}{\textbf{3,5-dbpy}} & Control       & $104 \pm 1$ & $87 \pm 8$ \\
    \cline{2-4}
                                       & Beam-on   & $104 \pm 2$ & $88 \pm 2$ \\
    \hline
    \hline
    \multirow{2}{*}{\textbf{2,6-dcpz}} & Control       & $170 \pm 1$ & $141 \pm 1$  \\
    \cline{2-4}
                                       & Beam-on   & $168 \pm 6$ & $142 \pm 5$  \\
    \hline
    \end{tabular}
    %}
    \label{tab:decay_t1s}
\end{table}

As can be seen in \cref{fig:decay_profiles}, for both the period before and after the beam is turned on there is no deviation between the beam-on runs and the control runs. This shows that there is no pronounced change to the decay rate in the presence of the beam. For \textbf{3,5-dcpy} the average $T_1$ values of the beam-on runs for before and after the beam were $167\pm 3$~s and $160\pm6$~s, 2\% shorter and no change from the control run respectively\footnote{Run 1 for \textbf{3,5-dcpy} had a beam current of \SI{5}{\nano\ampere}. All other runs had a beam current of \SI{10}{\nano\ampere}.}. For \textbf{3,5-dbpy} the average $T_1$ values of the beam-on runs before and after the beam were $104\pm2$~s and $88\pm2$~s, no change and 1\% longer than the control~run. For \textbf{2,6-dcpz} the average $T_1$ values of the beam-on runs before and after the beam were $168\pm6$~s and $142\pm5~$s, 1\% shorter and 1\% longer than the control run. Thus, for all three substrates, the $T_1$ values for the beam-on runs do not significantly deviate from the control runs, showing no evidence for any increased relaxation in the presence of the beam. A systematic effect, arising from the incorrect calibration of the larger pulse angles, slightly reduced the $T_1$ values in both the beam-on and control runs for the latter portion of the decay. This issue is discussed in more detail later.

\begin{figure}[htbp]
    \centering
    \includegraphics[width=1\columnwidth]{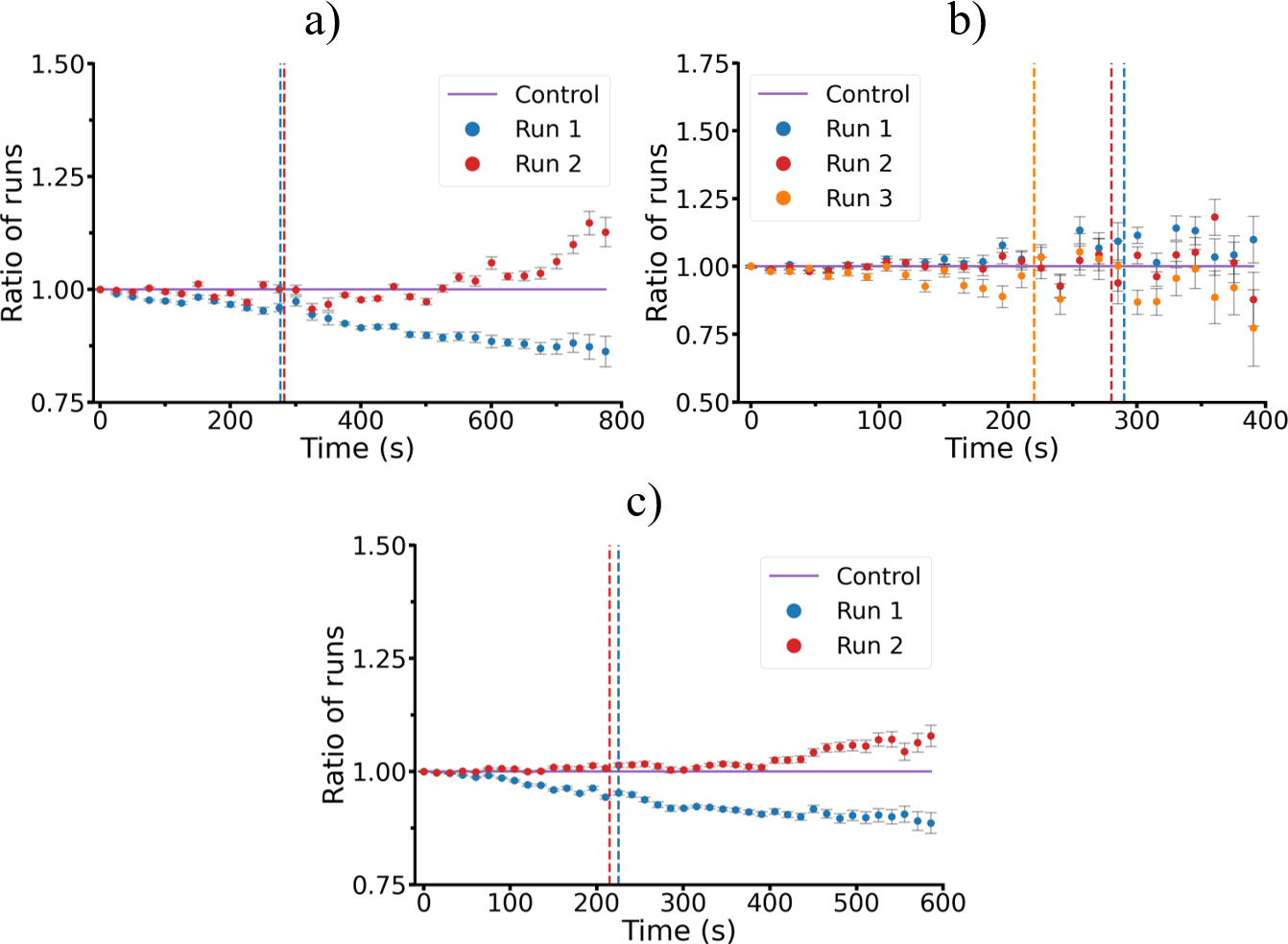}
    \caption{Ratio of the normalized polarization decay profiles for the beam-on runs to the control runs for a) \textbf{3,5-dcpy}, b) \textbf{3,5-dbpy},                and c)                \textbf{2,6-dcpz}. The dotted lines indicate the start time of the photon beam.}
    \label{fig:decay_ratios}
\end{figure}

To better display the deviation of the beam-on runs from the control runs, the ratios of their normalized decay profiles are shown in \cref{fig:decay_ratios}. At the time the beam was turned on, the average ratios were 0.98 ± 0.03 (\textbf{3,5-dcpy}), 1.03 ± 0.06 (\textbf{3,5-dbpy}), and 0.97 ± 0.05 (\textbf{2,6-dcpz}). At the end of the runs, the average ratios were 0.99 $\pm$ 0.19 (\textbf{3,5-dcpy}), 0.92 $\pm$ 0.17 (\textbf{3,5-dbpy}), and 0.98 $\pm$ 0.14 (\textbf{2,6-dcpz}), showing no significant deviation from unity. 

To further statistically assess whether the rate of relaxation of each sample changed in the presence of the beam, the instantaneous decay rates are compared between the beam-on and control runs. For this, a logarithmic transform is applied to the normalized decay profiles to linearize the exponential decay. The gradients of the newly transformed decay profiles, $y_{log}(t)$, are directly proportional to the decay rate of the polarization. The discrete gradients between consecutive data points are calculated as:
\begin{equation}
    G_n = \left( \frac{\Delta y_{log}(t)}{\Delta t} \right)_n = \frac{y_{log}(t_{n+1})-y_{log}(t_n)}{t_{n+1}-t_n},
\end{equation}
where $t$ is the time elapsed since the start of the decay measurement and $n$ is the scan number. The relative change in the decay rate in the beam over each interval is quantified by the ratio:

\begin{equation}
    R_n = \frac{G_{n,beam-on}}{G_{n,control}}.
\end{equation}

\begin{figure}[htbp]
    \centering
    \includegraphics[width=1\columnwidth]{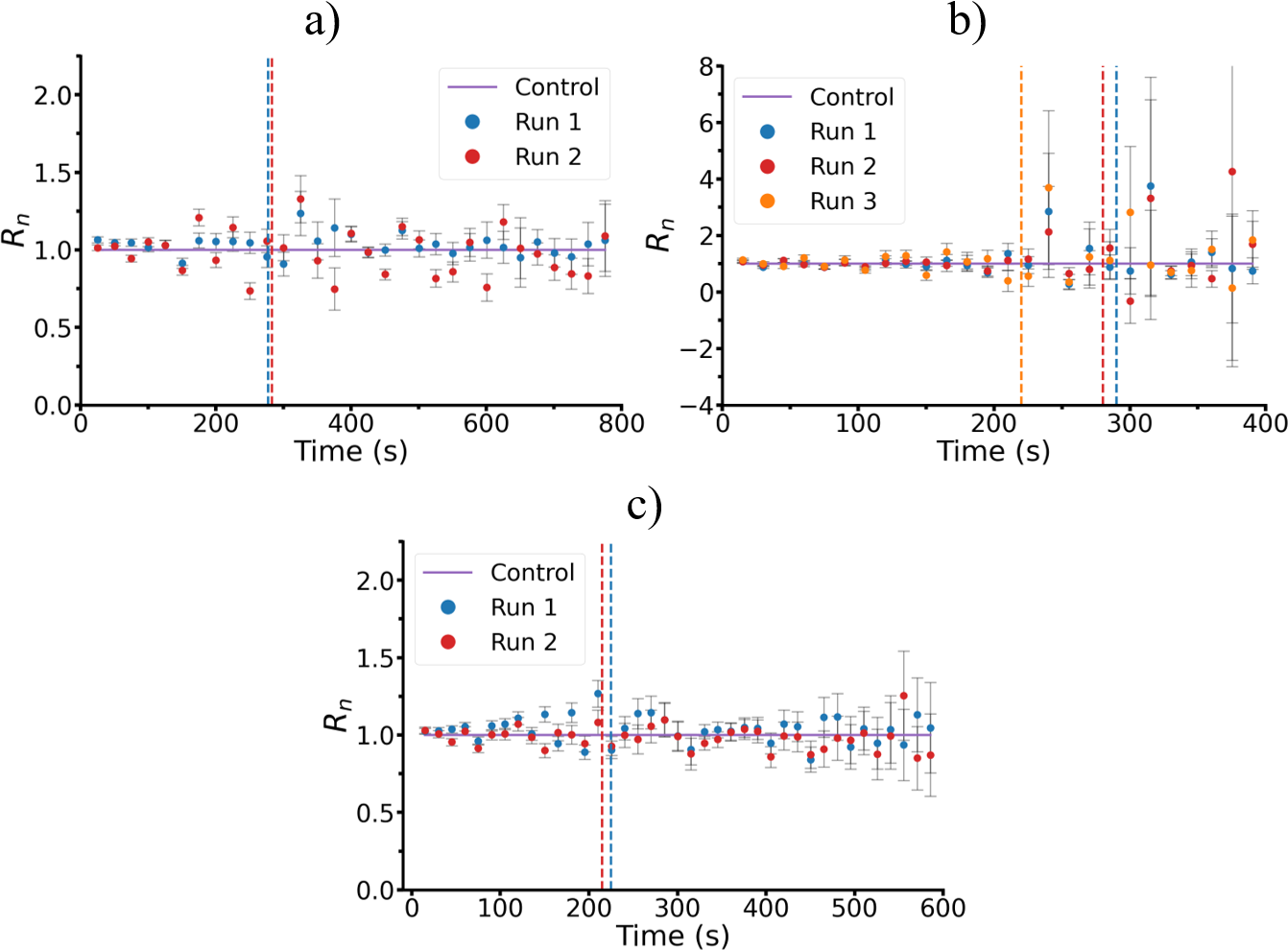}
    \caption{Ratios of log differentials, $R_n$, for a) \textbf{3,5-dcpy}, b) \textbf{3,5-dbpy}, and c) \textbf{2,6-dcpz}. The dotted lines indicate the start time of the photon beam.}
    \label{fig:log_diffs}
\end{figure}

Here, a value of $R_n$ of 2 would indicate that the polarization level during the beam-on run has decayed twice as fast as the control run over that time interval. The ratios $R_n$ will be referred to as the ratios of log differentials and are calculated for each beam-on run across all substrates, shown in \cref{fig:log_diffs}. The geometric mean and standard deviation of the ratios $R_n$, for before ($\mathbf{\overbar{R_{bef}}}$) and after ($\mathbf{\overbar{R_{aft}}}$) the beam for each substrate are presented in \cref{tab:rat_log_diffs}.

For \textbf{3,5-dcpy}, \textbf{3,5-dbpy}, and \textbf{2,6-dcpz}, $\mathbf{\overbar{R_{bef}}}$ is found to be $1.01 \divideontimes 1.10$, $0.99 \divideontimes 1.40$, and $1.02 \divideontimes 1.08$ respectively, all within error of unity. Here $\divideontimes$ denotes that the errors are multiplicative. $\mathbf{\overbar{R_{aft}}}$ is found to be $1.00 \divideontimes 1.13$, $1.10 \divideontimes 2.17$, and $1.00 \divideontimes 1.09$ for the three substrates respectively, again all within error of unity. This can be interpreted as the polarization levels of the \textbf{3,5-dcpy} and \textbf{2,6-dcpz} samples decaying at exactly the same rate as their respective control runs on average whilst the beam was incident, and not exceeding the decay rate of the control runs by more than 13\% and 9\% respectively to a confidence level of $1\sigma$. For the \textbf{3,5-dbpy} sample it is deemed that the error in $\mathbf{\overbar{R_{aft}}}$ is too large to pose sensible limits on the rate of decay in the beam.

\begin{table}[htbp]
    \centering
    \caption{Geometric mean and standard deviation of the ratios of log differentials before and after the beam.}
    \label{tab:rat_log_diffs}
    \begin{tabularx}{\columnwidth}{|>{\centering\arraybackslash}X|>{\centering\arraybackslash}X|>{\centering\arraybackslash}X|}
        \hline
        \textbf{Substrate} & $\mathbf{\overbar{R_{bef}}}$ & $\mathbf{\overbar{R_{aft}}}$ \\
        \hline
        \hline
        \textbf{3,5-dcpy}  & $1.01 \divideontimes 1.10$ & $1.00\divideontimes 1.13$ \\
        \hline
        \textbf{3,5-dbpy}  & $0.99 \divideontimes 1.40$ & $1.10\divideontimes 2.17$ \\
        \hline
        \textbf{2,6-dbpz}  & $1.02 \divideontimes 1.08$ & $1.00 \divideontimes 1.09$ \\
        \hline
    \end{tabularx}
\end{table}

\label{subsub:discussion_in_beam_pol_dec}
Overall, it is found that there is no evidence for an increased rate of depolarization for samples subject to the A2 photon beam. This was found for all three methods of analysis: the decay rate $T_1$, the ratios between the beam-on and control runs, and the analysis of point-to-point decay rates using the ratios of log differentials. The strictest limits on the rate of decay are found for the substrates \textbf{3,5-dcpy} and \textbf{2,6-dcpz}, whilst weaker limits are set for the substrate \textbf{3,5-dbpy}, due to the low SNR for this sample at later times.

It can be observed in \cref{fig:decay_profiles} that, for all substrates, the polarization decay profiles are not a smooth exponential decrease as expected, and contain small discontinuities and changes to $T_1$. These can be seen at at 375 s and 625 s in \Cref{fig:decay_profiles}a, 300 s in \Cref{fig:decay_profiles}b, and 300 s in \Cref{fig:decay_profiles}c. These features occur with a change in the tip angle in the measurement sequence, needed to address the decrease in signal over time, and are observed for both the beam-on runs and the control runs. They can thus be ruled out from being a beam-induced effect. The proposed explanation for this effect is a calibration error in the length of the RF pulse necessary to produce a specified tip angle in the sample. During the degassing procedure there was some loss of sample volume which led to the volume dropping slightly below that needed to fill the end region of the cell. If the sample volume within the MRI system was less than what was calibrated for, the tip angle produced by the RF pulse would be larger than requested. This would explain why the calculated $T_1$ values for both the beam-on and control runs were lower for the latter stages, as a higher proportion of the sample magnetization was being used up than had been calibrated for.

\subsection{High dose sample irradiation}
\label{subsec:high_dose_samp_irrad}
Here, we present results assessing the radiation damage within a SABRE sample after exposure to a high dose of \SI{3}{\kilo\gray}, attained in proximity to the electron beam dump within the A2 hall over four days of running. The sample chosen to be irradiated in this study was a replicate of the \textbf{3,5-dcpy} sample used for the in-beam tests.

A control measurement ran at York, designed to exclude the effects of sample aging, prepared a sample of identical composition and measured its achievable polarization and $T_1$ value before and after a four-day period, replicating the duration of the MAMI experiment. It was found that $T_1$ was unchanged, measured to be $153\pm2$s before and $153 \pm 4$s after the four-day duration. Additionally, the relative polarization after four days was 0.99 ± 0.06, showing no change to within the error margin.

The irradiated sample at MAMI was positioned vertically against the electron beam dump in the A2 hall, with a sample height of \SI{5}{\centi\metre}, a diameter of approximately \SI{1.1}{\centi\metre}, and a total sample volume of \SI{5}{\milli\litre}. The calculated electron flux through this position was \SI{20}{\mega\hertz}, leading to an estimate for the total electron flux using an estimated 80\% running time over four days to be \num{5.5e12} electrons, equivalent to that received from 15 minutes of irradiation at an electron beam current of \SI{1}{\nano\ampere}. Using an average energy deposition of \SI{23.6}{\mega\electronvolt\per{electron}}, found by Geant4 simulation, the total energy deposition was calculated to be \SI{20.9}{\joule}, with an absorbed dose of \SI{3.1}{\kilo\gray}.

\begin{figure}[htbp]
    \centering
    \includegraphics[width=1\columnwidth]{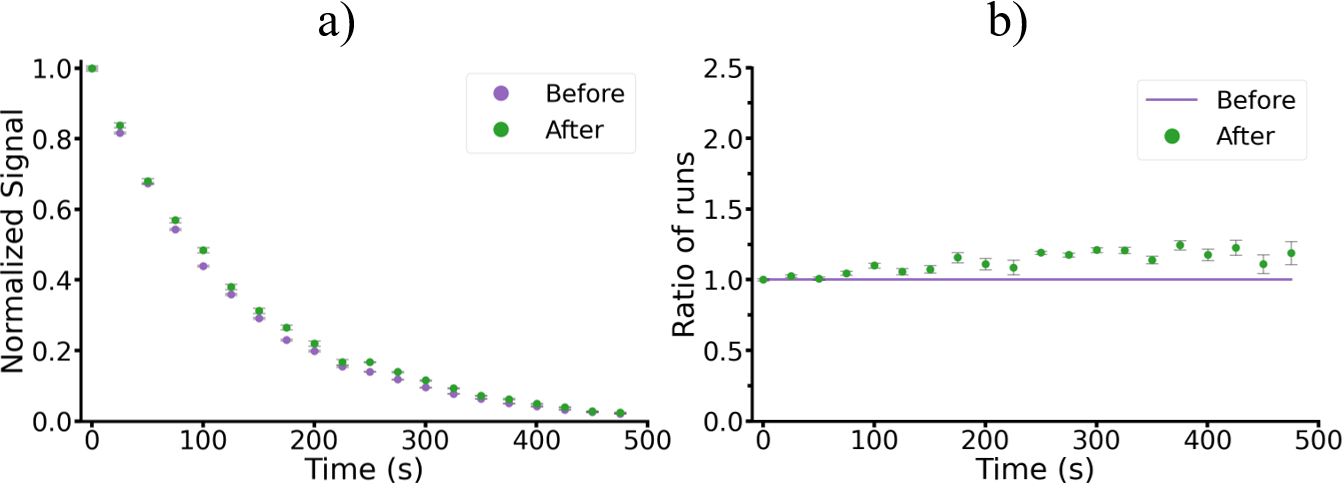}
    \caption{a) Normalized polarization decay profiles for before and after irradiation. b) Ratio of normalized decay profiles for before and after irradiation.}
    \label{fig:beam_dump_combined}
\end{figure}

The results of this experiment found a marginal decrease in the decay rate for the sample following irradiation, alongside a slightly reduced achievable polarization level, both shown in \cref{tab:irrad_t1_pol}. The polarization decay profiles for the sample before and after the irradiation process are shown in \cref{fig:beam_dump_combined}a. 

The fitted $T_1$ values were found to be 121 $\pm$ 1 s before and 126 $\pm$ 2 s after irradiation respectively, 4 $\pm$ 2\% higher after irradiation. Consistently, \cref{fig:beam_dump_combined}b shows the ratio of the post- to pre-irradiation decay profiles remaining above 1. Here the $T_1$ values for the irradiated sample were found to be shorter than for the control sample ran at York, attributed to a less thorough degassing procedure being possible at MAMI, resulting in a higher concentration of paramagnetic impurities such as oxygen. The lack of a decrease in the substrate $T_1$ after irradiation suggests that there was no significant accumulation of long-lived radicals within the sample.

The relative polarization level of the sample after irradiation was found to be 13\% lower than before, however, the inherent variability of the polarization technique meant that the standard error in the polarization levels for samples run at MAMI was $\pm$15\%, making this change not statistically significant. Absolute polarization levels could not be measured due to the insufficient sensitivity and resolution within the MRI system to acquire a thermal spectrum, and repeats to improve the statistical significance of this measurement were not possible due to time constraints. 

From these results, it is shown that there is no significant catalyst deterioration from the received dose. This experiment provides a first test of the radiation resilience of a SABRE sample, showing that a high fraction ($>$80\%) of the original polarization level is possible to be achieved post-irradiation, with no decrease to $T_1$, after a dose of around \SI{3}{\kilo\gray} was received.

\begin{table}[htbp]
    \centering
    \caption{$T_1$ and relative polarization values for the irradiated sample of \textbf{3,5-dcpy} before and after irradiation.}
    \begin{tabular}{|c|c|c|}
        \hline
         Sample & \textbf{\ce{^{1}H} \boldmath$T_1$ (s)} & Relative polarization \\
     \hline
    \hline
         Before irradiation & $121 \pm 1$ & 1 \\
    \hline
         After irradiation & $126 \pm 2$ & 0.87 \\
    \hline
    \end{tabular}
    
    \label{tab:irrad_t1_pol}
\end{table}

\subsection{Fluorescence measurements}
Initial measurements on fluorescent solutions containing SABRE material have shown that scintillation light can effectively propagate across a cell of length \SI{100}{\milli\metre} through solutions containing at least 50\% (v/v) of the SABRE substrate pyridine mixed with a liquid scintillator (Eljen technology EJ-309). At this concentration it was found that the average measured light output was only 34\% reduced compared to a pure liquid scintillator solution, despite the 50\% reduction in fluorescent material. Further results found that the addition of small quantities (\SI{0.16}{\milli\mol\per\litre}) of the SABRE precatalyst [Ir(IMes)(COD)Cl] to a solution of liquid scintillator with 30\% (v/v) pyridine resulted in a 43\% reduction in the measured light output. Upon activation of the precatalyst to form the active SABRE catalyst, \ce{[Ir(IMes)(H)_2(pyridine)_3]Cl}, it was found that this reduction in light output became only 23\% when compared to the liquid scintillator and pyridine solution. These initial results are promising for the production of polarized active detector media using SABRE.

\section{Conclusions}
\label{sec:conclusions}
In this work, the first in-beam measurements of depolarization effects on SABRE-polarized material are presented. These measurements were made for the SABRE substrates \textbf{3,5-dcpy}, \textbf{3,5-dbpy}, and \textbf{2,6-dcpz} and were achieved at the A2 facility at MAMI with the use of a portable benchtop MRI system positioned in line with the photon beam. Additionally, measurements were made on the effects of a high radiation dose, attained in proximity to the MAMI electron beam, on the polarization levels and decay rate of a SABRE sample. 

These results place an upper limit on the influence on the decay rate of SABRE-polarized material due to exposure to a photon beam with the aforementioned properties. The strongest constraints are obtained for the substrates \textbf{3,5-dcpy} and \textbf{2,6-dcpz}, whilst weaker limits are imposed for the substrate \textbf{3,5-dbpy} for which the signal strength at later times was decreased. This study finds no significant increase in the rate of polarization decay within the photon beam for any of the samples. Consequently, there is no evidence to show any substrate dependence of the effects of beam-induced depolarization. Current polarized target experiments with photon beams use intensities up to an order of magnitude higher than were possible for this investigation\footnote{Higher untagged photon beam intensities could be achieved at the A2 facility by utilizing either a thicker radiator or a higher electron beam intensity of up to 100 nA.} \cite{shepherd2009gluex}, and thus further testing is needed in order to show that SABRE can operate as the polarization method for photon beams of such intensity. The results here, however, show promise.

A high level of polarizability was retained for the sample subject to a strong radiation dose of around \SI{3}{\kilo\gray}, with no measurable detriment to the post-irradiation $T_1$ value. This key result addresses a significant challenge faced by traditional polarized target technologies at high beam intensities, radiation damage of the target material. It demonstrates the potential for SABRE-polarized material to operate within $\geq$\SI{10}{\nano\ampere} electron beams with continuous replenishment of the target material, possible only due to the solution-state nature of SABRE.

Specifically, for the substrates \textbf{3,5-dcpy} and \textbf{2,6-dcpz}, which exhibited a $T_1$ of ${\sim}$\SI{160}{\second} in this study, a fractional replacement rate of $\geq$ \SI{0.06}{\per\second} would be needed to keep the polarization level of a target above $90\%$ of its nominal value if polarized outside of the beam. At this replacement rate, such a target could potentially handle an electron beam current of \SI{50}{\nano\ampere} without appreciable sample degradation, extrapolating the result from this study.

Liquid targets, such as those produced by SABRE, are self-repairing, where convection effectively dissipates local beam heating and radiation damage effects. This, alongside short replacement times of the target material, mitigates radiation damage without the need for experimental downtime. Combined with rapid polarization build-up (seconds to minutes) and low implementation costs, SABRE is positioned as a highly promising candidate for next-generation polarized target applications. Furthermore, the compatibility of SABRE-polarized targets with  $\unit{\milli\tesla}$-strength fields establishes it as one of the few technologies capable of facilitating both $4\pi$ angular acceptance within a detector setup and rapid field reversal — the latter allowing for the control of systematic errors within an experiment.

Despite the advantages that SABRE brings over traditional polarized target technologies, challenges remain. The foremost of which are scaling up the volumes of polarized material from \unit{\micro\litre} to \unit{\milli\litre} scales, alongside optimizing the dilution factor of the polarized material. It is important for future work to extend the limits of heat and radiation deposition from particle beams on SABRE samples, with high intensity electron beams, such as the X1 facility at MAMI, being a clear next step. In general, there are challenges in using SABRE targets with electron beams which, due to their strong ionization, are transported in evacuated beam pipes. Future work in York will explore the challenges in introducing polarized SABRE materials into vacuum.

\section*{Declaration of competing interests}
The authors declare that they have no known competing financial interests.

\section*{Acknowledgements}

The authors would like to acknowledge Dr Victoria Annis for the synthesis of the [Ir(IMes-\textit{\ce{d_22}})(COD)Cl] precatalyst. Additionally we would like to express our gratitude for the backing provided by the MAMI facility and the A2 collaboration at Johannes Gutenberg University Mainz, alongside the travel support received via the European Union's Horizon 2020 research and innovation program under grant agreement No. 824093. This work was supported by the UK Science and Technology Facilities Council (STFC) grant ST/W004852/1. This research was also supported in part by the Cluster of Excellence ``Precision Physics, Fundamental Interactions, and Structure of Matter'' (PRISMA++ EXC 2118/2) funded by the German Research Foundation (DFG) within the German Excellence Strategy (Project ID 390831469).

\section*{Data availability}
Data are available upon request.
\bibliographystyle{elsarticle-num} 
\bibliography{references}



%% For example:

%% \bibitem[Aladro et al.(2015)]{Aladro15} Aladro, R., Martín, S., Riquelme, D., et al. 2015, \aas, 579, A101

%%\end{thebibliography}

\end{document}